\documentclass[12pt]{article}
\pdfoutput=1
\usepackage{microtype}
\usepackage{sectsty}
\usepackage{mathrsfs}
\usepackage{amsmath}
\allsectionsfont{\sffamily}
\usepackage[nosort]{cite}

\usepackage[pdftex]{graphicx}
\usepackage[font={scriptsize,singlespacing}]{caption}
\usepackage{epstopdf}

\usepackage{amsmath}
\usepackage{amssymb}
\usepackage{subfigure}
\usepackage{hyperref}
\usepackage{url}
\usepackage{xcolor}
\usepackage{color}
\definecolor{amaranth}{rgb}{0.9, 0.17, 0.31}
\definecolor{purple(munsell)}{rgb}{0.62, 0.0, 0.77}
\definecolor{americanrose}{rgb}{1.0, 0.01, 0.24}
\definecolor{palatinateblue}{rgb}{0.15, 0.23, 0.89}
\definecolor{royalblue(web)}{rgb}{0.25, 0.41, 0.88}
\definecolor{hanpurple}{rgb}{0.32, 0.09, 0.98}
\definecolor{beaublue}{rgb}{0.74, 0.83, 0.9}
\definecolor{carminered}{rgb}{1.0, 0.0, 0.22}
\definecolor{brightpink}{rgb}{1.0, 0.0, 0.5}
\definecolor{vividviolet}{rgb}{0.62, 0.0, 1.0}
\hypersetup{ linktoc=all,
    colorlinks, linkcolor={palatinateblue},
    citecolor={brightpink}, urlcolor={amaranth}}

\newcommand{\changeurlcolor}[1]{\hypersetup{urlcolor=#1}}    
\def\sideremark#1{\ifvmode\leavevmode\fi\vadjust{\vbox to0pt{\vss
 \hbox to 0pt{\hskip\hsize\hskip1em
 \vbox{\hsize2cm\tiny\raggedright\pretolerance10000
 \noindent #1\hfill}\hss}\vbox to8pt{\vfil}\vss}}}%
\newcommand{\bo}{\raise-1mm\hbox{\Large$\Box$}}

\newcommand{\be}{\begin{equation}}
\newcommand{\ee}{\end{equation}}
\newcommand{\bea}{\begin{eqnarray}}
\newcommand{\eea}{\end{eqnarray}}

\renewcommand{\d}[1]{\ensuremath{\operatorname{d}\!{#1}}}

\begin{document}
\thispagestyle{empty}
\begin{center}

\null \vskip-1truecm \vskip2truecm

{\LARGE{\bf \textsf{Giant Tortoise Coordinate}}}
\textrm\textrm
\vskip1truecm
\textbf{\textsf{ Michael R.R. Good$^{*}$, Yen Chin Ong$^{\dagger}$\footnote{Corresponding Author: ycong@yzu.edu.cn.}, \\Aizhan Myrzakul$^{*}$ and Khalykbek Yelshibekov$^{*}$}}\\{\footnotesize\textsf{$^{\dagger}$Center for Gravitation and Cosmology, College of Physical Science and Technology,\\ Yangzhou University, Yangzhou, China}}\\ 
{\footnotesize\textsf{$^{*}$ Department of Physics, School of Science and Technology,\\ Nazarbayev University, Astana,  Kazakhstan}}\\



\end{center}
\vskip1truecm \centerline{\textsf{ABSTRACT}} \baselineskip=15pt

\medskip
The giant tortoise coordinate is a moving mirror inspired generalization of the Regge-Wheeler counterpart that demonstrates a unitary evaporating black hole emitting a total finite energy.

\vskip0.4truecm
\hrule

\section{Collapse Geometry and the Moving Mirror Model}\label{sec:metric} 
The moving mirror is a perfectly reflecting boundary, accelerating according to some trajectory. Qualitatively, the mirror disturbs the quantum field and reflects virtual particles into real particles. 
In this paper we briefly review the exact correspondence of the collapse of a null shell using the tortoise coordinate to that of the black mirror \cite{Good:2016oey}. It was shown that the black mirror emits an infinite amount of energy \cite{Good:2016atu}, which is arguably unphysical. There it was also shown that an improved model -- the ``drifting black mirror" -- emits a finite amount of energy, as anticipated by Wilczek \cite{Wilczek:1993jn}. We demonstrate that the corresponding tortoise coordinate for finite energy emission is the giant tortoise coordinate. 

Imposing suitable boundary conditions at the origin of coordinates in the Hawking effect \cite{Hawking:1974sw} is explicitly shown to be a moving mirror \cite{Davies:1976hi, Davies:1977yv}.  For scope of the mirror's basic properties see \cite{Good:2016bsq, Anderson:2015iga, Good:2015jwa, Hotta:1994ha, Good:2013lca, Good:2012cp}.  The mirror corresponds to the origin $r=0$ of the Schwarzschild coordinates, i.e. the singularity of the black hole (the ``center" of the black hole), \textit{not} the event horizon.

The geometric distortion of spacetime during collapse has the same effect as a rapidly receding mirror.  The severe red-shifted rays that avoid being trapped behind the incipient event horizon are the same red-shifted rays that are reflected off the mirror.  The new model presented here does not let the black hole origin escape to null infinity, i.e. the mirror recedes to time-like infinity.  In the usual case, the radial distance from the point of view of wave propagation is measured in intervals of the tortoise coordinate $r^*$ which diverges to $-\infty$ at the black hole horizon.  Instead we utilize the giant tortoise coordinate which does not diverge to $-\infty$ at the black hole horizon.  

In the giant tortoise case all rays escape the black hole (intersect the mirror).  An initial pure state can evolve to look thermal for an arbitrary long time but no information is ever lost.  This is because the model does not represent a complete black hole evaporation, but that of a remnant. The entire state outside and inside the remnant, taken together, remains pure at all time.  The red-shifting of the field modes is eternal despite the absence of radiative evaporation signaled by an asymptotically zero stress-energy flux. The physical explanation is that the field modes are disturbed by the non-evaporative remains of the black hole, i.e. the remnant is still capable of red-shifting incoming modes. This might seem strange since one expects remnants to be microscopic, but curved spacetime geometries allow remnants to have non-trivial large interiors that could red-shift incoming modes \cite{HL,COY}.

\section{Black Hole-Black Mirror}
In this section we derive the trajectory of the black mirror from the tortoise coordinate and we show the matching of the metrics over the shock wave, which is the intermediate collapse region shrunk down to, ideally, a single null surface located at some $v = v_0$ as described in, for instance, \cite{Fabbri:2005mw}.

\subsection{The Tortoise and the Mirror}
The usual tortoise coordinate or the Regge-Wheeler $r^*$ is defined as:
\begin{equation}
r^*\equiv r+2M \ln\left(\frac{r}{2M}-1\right), \label{tortoise}
\end{equation}
with the well-known corresponding derivative useful for Eddington-Finkelstein coordinates:
\be \frac{\d r^*}{\d r} = \frac{1}{1-2M/r} \equiv f^{-1}. \label{dtortoise} \ee
The usual spacetime matching procedure for collapse is given in a nice treatment in Fabbri (2005) \cite{Fabbri:2005mw}, which we briefly review here.  To solve for the red-shift function, $u_{\textrm{out}}$, first one matches inside and outside of the shock wave, 
\begin{equation}
r(v_0, u_{\textrm{in}})=r(v_0, u_{\textrm{out}}), 
\end{equation}where
\begin{equation}
r(v_0, u_{\textrm{in}})=\frac{v_0-u_{\textrm{in}}}{2}\quad \quad 
\text{and} \quad\quad
r^*(v_0, u_{\textrm{out}})=\frac{v_0-u_{\textrm{out}}}{2}.\label{4}
\end{equation} 
Thus, using the tortoise coordinate, Eq.~(\ref{tortoise}),
\be r(v_0,u_{\textrm{out}}) + 2M \ln\left(\frac{r(v_0,u_{\textrm{out}})}{2M}-1\right) = \frac{v_0 - u_{\textrm{out}}}{2}.\label{matched}\ee
Eqs.~(\ref{tortoise})-(\ref{matched}) lead to the matching solution for the Eddington-Finkelstein metric to flat geometry at the shock wave for a black hole with an event horizon:

\begin{equation}
u_{\textrm{out}}=u_{\textrm{in}}-4M \ln\frac{|v_H-u_{\textrm{in}}|}{4M},  \label{m1}
\end{equation}
where $v_H\equiv v_0-4M$ is a null ray location which at $u_{\textrm{out}}=+\infty$ forms the event horizon. 

One can see that the mirror ray-tracing function in \cite{Good:2016oey}, $f(v)$, is the same as the black hole matching condition, $u_{\textrm{out}}(u_{\textrm{in}})$, that is, $f(v) \leftrightarrow u_{\textrm{out}}(u_{\textrm{in}})$.  We substitute $ u_{\textrm{out}} \equiv t(x) - x$ and $u_{\textrm{in}} \equiv t(x) + x$, into Eq.~(\ref{m1}), 
and solve for $t(x)$,
\be t(x) = v_H - x - 4M e^{x/2M}.\label{omex}\ee
This is the time-space trajectory of the moving mirror\footnote{In \cite{Good:2016oey}, the mirror is called Omex for short, due to the Omega constant, $\Omega e^{\Omega} = 1$, and exponent argument.}, with $\kappa = 1/(4M)$, as investigated in Good (2016) \cite{Good:2016oey}.  The worldline of the origin is the moving mirror, where  its “motion” is an effective representation of the distortion of spacetime in
the collapse geometry.  In 3+1 dimensions the coordinates are from $0<r<\infty$ and $-\infty < x < \infty$.  The connection now between the moving mirror trajectory, Eq.~(\ref{omex}), and the black hole matching condition, Eq.~(\ref{m1}), is explicit and analytic.

\subsection{Spacetime Matching}
Following Unruh (1976) \cite{Unruh:1976db}, Wilczek (1993) \cite{Wilczek:1993jn}, Massar (1996) \cite{Massar:1996tx}, and Fabbri (2005) \cite{Fabbri:2005mw}, we outline a sketch of the null shell collapse. We work in 1+1 dimensions for clarity, however, the arguments readily generalize to 3+1 dimensions.  Adopting mostly the notation of Wilczek, the interior and exterior regions of a spacetime are given as:

\begin{equation}\label{cg}
\d s^2=
\begin{cases}
-\d t_{\textrm{in}}^2+\d r^2,~~~~~~~~~~\text{for}~t_{\textrm{in}}+r\leqslant v_{\textrm{in},0}, \\
-f\d t_{\textrm{out}}^2+f^{-1}\d r^2,~~\text{for}~t_{\textrm{out}}+r> v_{\textrm{out},0}, 
\end{cases} 
\end{equation}
where $f=1-{2M}/{r}$. Now we would like to express each of these metrics in a static form using lightcone coordinates. We have
$$
\begin{cases}
u_{\textrm{in}}=t_{\textrm{in}}-r\\
v_{\textrm{in}}=t_{\textrm{in}}+r
\end{cases},~\text{for~the~inner~region};
$$
$$
\begin{cases}
u_{\textrm{out}}=t_{\textrm{out}}-r^* \\
v_{\textrm{out}}=t_{\textrm{out}}+r^*
\end{cases},~\text{for the outer region}.
$$
Using these and Eq.~(\ref{dtortoise}), the spacetime metrics (\ref{cg}) can be rewritten in null coordinates as:
\begin{equation}\label{cg1}
\d s^2=
\begin{cases}
-\d u_{\textrm{in}}\d v_{\textrm{in}},~~~~~~\text{for}~v_{\textrm{in}} \leqslant v_{\textrm{in},0}, \\
-f\d u_{\textrm{out}}\d v_{\textrm{out}},~~\text{for}~v_{\textrm{out}}
> v_{\textrm{out},0},
\end{cases}
\end{equation}
Let us define both regions in the $u_{\textrm{out}}$-$v_{\textrm{out}}$-coordinates, i.e., we seek expressions for $u_{\textrm{in}}(u_{\textrm{out}})$ and $v_{\textrm{in}}(v_{\textrm{out}})$. As the spacetime is flat at the early stages of the collapse, the coordinate systems for both regions are the same.  This allows for the identification of $v_{\textrm{in}}(v_{\textrm{out}})=v_{\textrm{out}}$.  In both systems, $r$ should agree, which implies the implicit relation along the shell worldline, $v=v_{0}$, giving
\begin{equation}
r^*\left(r=\frac{v_{0}-u_{\textrm{in}}(u_{\textrm{out}})}{2}\right)=\frac{v_0-u_{\textrm{out}}}{2},\label{tortoise1}
\end{equation}
which, in turn, along with the derivative of Eq.~(\ref{dtortoise}) yields:
\begin{equation}
\frac{\d u_{\textrm{in}}}{\d u_{\textrm{out}}}=f(u_{\textrm{out}},v_0), \label{derivative}
\end{equation}
and the metric becomes:
\begin{equation}\label{out-out}
\d s^2=
\begin{cases}
-f(u_\textrm{out},v_0) \d u_{\textrm{out}}\d v_{\textrm{out}},~~~~\text{for}~v_{\textrm{out}}\leqslant v_0\\
-f(u_\textrm{out},v_\textrm{out})\d u_{\textrm{out}}\d v_{\textrm{out}},~~\text{for}~v_{\textrm{out}}>v_0,
\end{cases}
\end{equation}
which is continuous along $v_0$. It is worth to note that this metric holds if $r$ is positive in the 3+1 dimensional case, i.e. when $v_\textrm{out}\geqslant u_\textrm{in}(u_\textrm{out})$ (which is robust in the 1+1 dimensional case too, where a static mirror can be placed at $r=0$). Thus the worldline origin is defined as:
\begin{equation}
v_s(u_\textrm{out})=u_\textrm{in}(u_\textrm{out}). \label{worigin}
\end{equation}
The origin in the 1+1 dimensional case allows no field modes to pass beyond it, and in the 3+1 case, since $r$ is non-negative, the origin behaves identically as a perfectly reflecting mirror. 
Let us define the global spacetime in $u_{\textrm{in}}$-$v_{\textrm{in}}$-coordinates:
\begin{equation}\label{in-in}
\d s^2=
\begin{cases}
-\d u_{\textrm{in}}\d v_{\textrm{in}},~~~~~~~~~~~~~~~~~~~~~~~~~~~~~~~~~\text{for}~v_{\textrm{out}}\leqslant v_0, \\
-f(u_\textrm{out},v_\textrm{out})f^{-1}(u_\textrm{out},v_0)\d u_{\textrm{in}}\d v_{\textrm{in}},~~\text{for}~v_{\textrm{out}}> v_0.
\end{cases}
\end{equation}
This metric is analytic and single-valued on the horizon, with $f = 0$.  The origin remains stationary at $u_{\textrm{in}} = v_{\textrm{in}}$ until the shell arrives at it.  Taking into account the form of the function $f$ and the derivative (\ref{dtortoise}) gives our usual tortoise coordinate, Eq.~(\ref{tortoise}) with an integration constant:
\begin{equation} \label{tortoise2}
r^*(r)=r+2M\ln |r-2M|+c,
\end{equation}
which leads to the matching solution of the Eddington-Finkelstein metric:
\begin{equation}\label{match}
u_{\textrm{out}}=u_{\textrm{in}}-4M\ln\left\vert\frac{-4M-u_{\textrm{in}}+v_{0}}{2}\right\vert-2c,
\end{equation}
where $c=-2M\ln 2M$, so that the matching condition agrees with Eq.~(\ref{m1}), but is ultimately arbitrary and does not change the physics.  An exact inversion of Eq.~(\ref{match}) is possible \cite{Good:2016oey,Good:2016atu} without approximation\footnote{An inversion of the tortoise coordinate itself, Eq.~(\ref{tortoise}), is analytically tractable and the result is in terms of the Lambert W-function \cite{Boonserm:2008zg}. }.  The radiation resulting from this matching condition has sufficient justification from the identification of particle and energy production to conclude that the collapse geometry is precisely modeled by a particular moving mirror: ``the black mirror'' of Eq.~(\ref{omex}). The mirror in this model (and in our modified model in the next section) emerges at the origin of coordinates, in which the interior geometry is flat, with usual coordinates $(r,t_\text{in}) \rightarrow (r,u_\text{in})$:
\be \d s_\text{in}^2 = -\d t^2_\text{in}+\d r^2 = -\d u^2_\text{in} - 2 \d u_\text{in} \d r. \label{outFINK}\ee
The null shell will follow the geodesic $v=v_0$ where the event horizon is $v_H \equiv v_0 - 4M$. The exterior metric is
\be \d s^2_\text{out} =  -f \d t^2_\text{out}  + f^{-1} \d r^2 =- f \d u^2_\text{out} - 2 \d u_\text{out} \d r,\ee
for the exterior geometry of a thin shell of matter. 
 This is the usual Eddington-Finkelstein outgoing metric, which will be altered in the following section.

\section{Black Hole Corresponding to Drifting Mirror} 
\subsection{The Drifting Mirror and the Giant Tortoise}
In order to construct a history for finite total energy emission, we restrict the black mirror to speeds less than the speed of light (asymptotic speeds as well).  This amounts to restricting the regularity condition in the black hole case to travel at sub-light speeds, even asymptotically.  The position of the center of the black hole cannot, in any coordinate system, diverge.  This gives a new time-space trajectory \cite{Good:2016atu},
\be t(x) \quad \longrightarrow \quad t(x,\xi) = v_H - \frac{x}{\xi} - 4 M e^{\frac{x}{2M\xi}}, \label{domex}\ee
where $0 < \xi < 1$ is the asymptotic drifting speed. The equivalent matching condition is simply the associated ray-tracing function,
\be u_{\textrm{out}} = u_{\textrm{in}} - 4 M \xi \ln\left[\frac{1-\xi}{2}\mathcal{W}\left(\frac{2 e^{\frac{ v_H - u_{\textrm{in}}}{2 M (1-\xi)}}}{1-\xi}\right)\right], \label{m2} \ee
where $\mathcal{W}$ is the product log, and $v_H$ would normally be the position of the event horizon.  We call $v_H$ the position of the ``residual horizon'' using the analogy of the removal of the acceleration horizon in the mirror case which left a residual acceleration horizon. In addition, $\xi$ now acts like a singularity removal parameter. Here, by singularity we do not mean the black hole central singularity, but rather the aforementioned divergence. These matching solutions $u_{\textrm{out}}$, Eq.~(\ref{m1}) and Eq.~(\ref{m2}), are arbitrarily equivalent for $\xi \to 1$. 
This also means that the corresponding black hole has no event horizon in the strict sense, one can get an event horizon arbitrarily close to forming, and thus for any ``practical'' purpose it behaves like a black hole\footnote{It might be interesting to check the behavior of the apparent horizon, if any, during the collapse. The question of whether a horizon (of any sort) is formed during gravitational collapse when quantum effects are taken into account has recently received renewed attention, see, e.g., \cite{1710.01533,Mann:2018jcf}.}.

\begin{figure}[ht]
\begin{center}
\mbox{\subfigure{\includegraphics[width=3.0in]{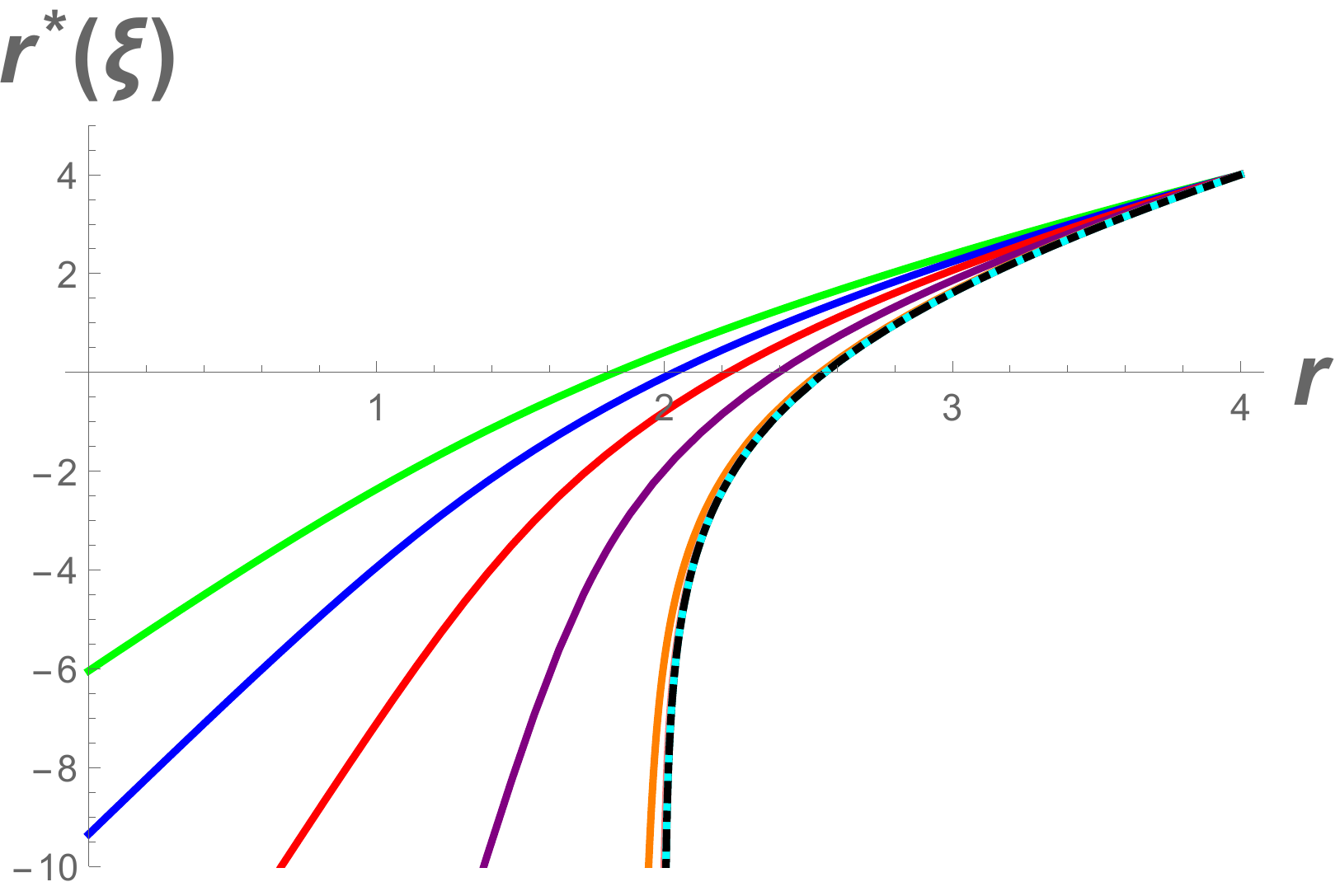}}\quad
\subfigure{\includegraphics[width=3.0in]{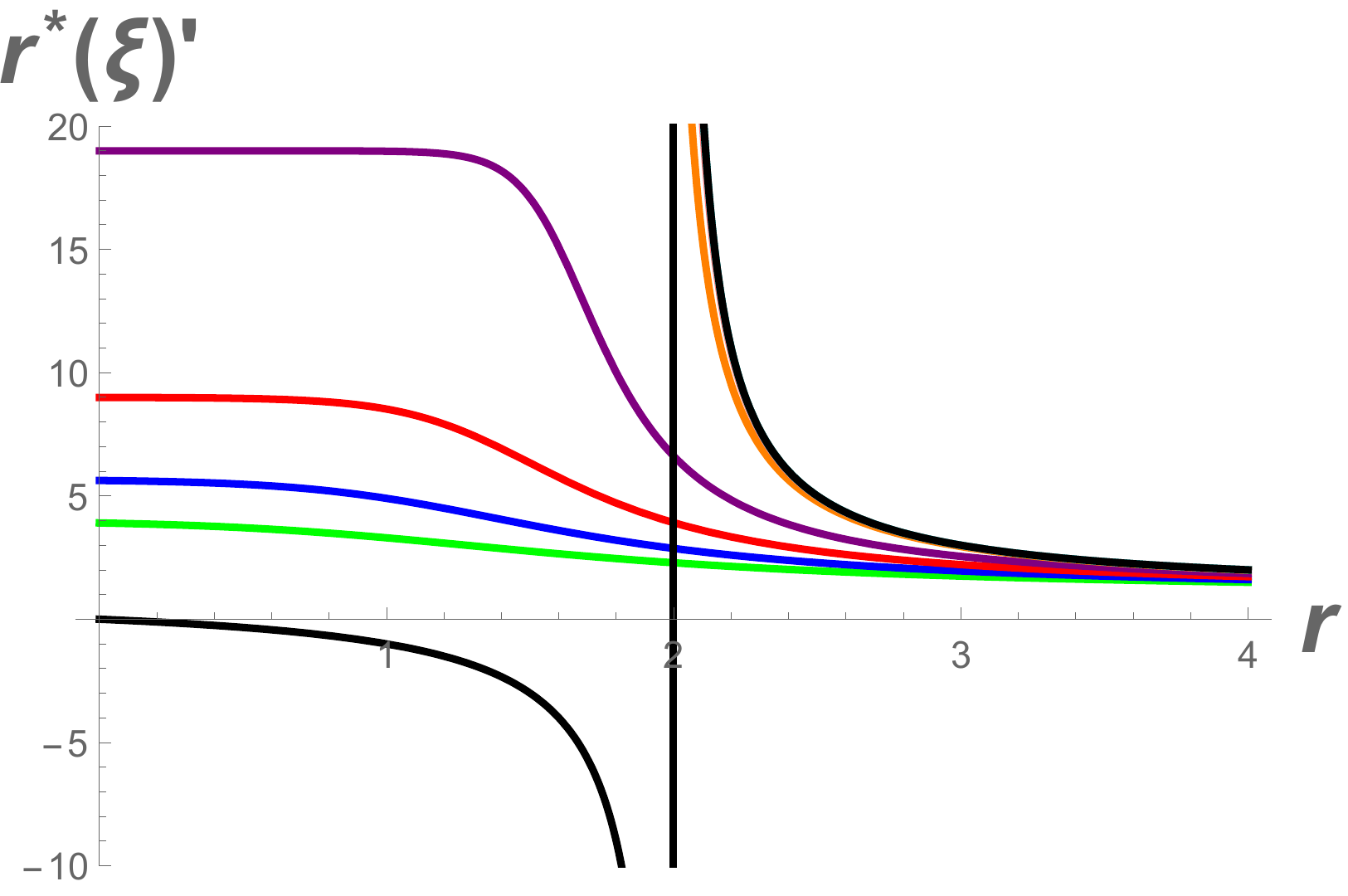} }}
\caption{\label{fig:tort} The giant tortoise coordinate and its derivative, with various $\xi$ values. The black lines are the usual Regge-Wheeler tortoise coordinate and its derivative.  The faster the final drifting speed, the more closely the generalized coordinate becomes to the usual tortoise coordinate. Green, Blue, Red, Purple are $\xi = 0.6, 0.7, 0.8, 0.9$, respectively. Orange, Pink, and Cyan are nearly piled up on top of the tortoise (Black) with $\xi = 0.99, 0.999, 0.99995$, respectively. } 
\end{center}
\end{figure}

The world line of the origin is given by Eq.~(\ref{m2}) in both the 1+1 D and 3+1 D cases.  Since nothing can go beyond the regular origin to negative $r$ in 3+1 dimensions, it illustrates the trajectory of the perfectly reflecting moving mirror. In the 1+1 D case it represents the dynamic origin of coordinates where the field is zero (due to the imposition of a static mirror in $(t_{\textrm{in}},r)$ coordinates). 
Therefore, it is a simple task to use the generalized matching solution Eq.~(\ref{m2}) to find what we called the giant tortoise coordinate,


\be \label{giant} \boxed{r^*(\xi) \equiv r+ 2 M \xi  \ln \left[\frac{1-\xi}{2} \mathcal{W}\left(\frac{2 e^{\frac{r-2 M}{M (1-\xi)}}}{1-\xi}\right)\right]},\ee
named ``giant'' due to its generalized form relative to the usual tortoise coordinate, Eq.~(\ref{tortoise}). {Eq.~(\ref{giant}) is the key mathematical result of this paper.}  The two tortoise coordinates, $r^*$ and $r^*(\xi)$, have the same limiting behavior as long as $\xi \approx 1$.  Strictly, there is no singularity in Eq.~(\ref{giant}) at $r = 2M$ as there is in Eq.~(\ref{tortoise}), as long as $\xi \neq 1$: $r^*(\xi)_{r=2M} = 2M \left(1- \xi \mathcal{W}(\sigma)\right)$. Here $\sigma \equiv 2/(1-\xi)$. An arbitrary close approach $\xi \rightarrow 1$ is allowed. See Fig.~(\ref{fig:tort}) for a comparison between the tortoise coordinate and the giant tortoise coordinate.
\subsection{Sewing Spacetime}
The generalized spacetime matching with the giant tortoise coordinate is \textit{not} strictly continuous.  It is, however, arbitrarily close.  The outgoing metric of Eq.~(\ref{outFINK}) becomes 
\be \d s^2 = -f \d u^2_\text{out}  -2 P \d u_\text{out} \d r + Q \d r^2. \label{newmetric} \ee
The functions $P\equiv P(r,M,\xi)$ and $Q\equiv Q(r,M,\xi)$ are defined as follows:
\be P \equiv f \frac{\d r^*(\xi)}{\d r}, \quad Q \equiv f^{-1}(1- P^2).\ee
The derivative of the giant tortoise coordinate is
\be \label{dtort2} \frac{\d r^*(\xi)}{\d r} = \frac{\sigma -2}{W\left(\sigma  e^{\left(r/2M-1\right) \sigma }\right)+1}+1. \ee
This expression and the derivative of the usual tortoise coordinate, Eq.~(\ref{dtortoise}), are the same as long as $\sigma \gg 2$ at some finite value of $\sigma$.  The requirement that $\sigma \neq \infty$ is ensured since strictly $\xi \neq 1$.  That is, for $\xi \approx 1$ the derivative of the giant tortoise coordinate is
\be \frac{\d r^*(\xi)}{\d r} \approx f^{-1}. \ee
In the limit $\xi \rightarrow 1$, the new generalized metric recovers the Eddington-Finkelstein metric because $P\rightarrow 1$ and $Q\rightarrow 0$. This new generalized metric of Eq.~(\ref{newmetric}) is of particular interest for its exact correspondence as the black hole analog of the drifting black mirror \cite{Good:2016atu}.  

While it is clear there is no exact continuity of $r$ due to the strict requirement that $P\neq 1$ and $Q\neq 0$, a continuity approximation can be made to be as arbitrarily close as one requires.  The relative validity of the approximation $P \approx 1$ and $Q\approx 0$ as increased continuity across $r$ of the shell is directly related to the increased total finite energy emitted by the evaporation process.   

\section{Finite Evaporation Energy}
A primary advantage of the use of the giant tortoise coordinate is the finite evaporation energy that is globally emitted during the collapse history.  The energy is consistent from both the beta Bogolubov coefficient calculation via quanta summing and from the stress-tensor (as can also be verified numerically \cite{Good:2013lca} via wavepackets \cite{Hawking:1974sw}).  The finite evaporation energy found from the stress-energy has a simpler derivation, see \cite{Good:2016atu}, and is:
\be E = \frac{\kappa}{24\pi}\left(\gamma^2 + \frac{\eta}{\xi}\right) \approx  \frac{\kappa}{24\pi}\gamma^2, \label{energy} \ee 
where $\kappa = 1/(4M)$ is the surface gravity, $\gamma \equiv 1/\sqrt{1-\xi^2}$ is the final drifting Lorentz factor, $\eta \equiv \tanh^{-1}\xi$ is the final drifting rapidity, and $\xi<1$ is the final drifting speed.  The expansion on the right is found by assuming very high drift speeds, $\xi \approx 1$.  
\subsection{Temperature and Speed}
With two assumptions (1) global energy of the spacetime is conserved, $M = E(\xi) + m$ (which is not a needed assumption for the energy Eq.~(\ref{energy}) to be finite), and (2) the mass of the remnant, $m$, is small compared to the mass of the black hole, $M\gg m$, where the black hole emits thermal radiation for some extended time, $\gamma\gg 1$, then
\be \gamma^2 = 96 \pi M^2. \label{speedmass}\ee
This gives a relationship between the temperature, $T= \kappa/(2\pi)$, and the drift speed, $\xi$, of the mirror (origin of coordinates):
\be T = \sqrt{\frac{6}{\pi}}\epsilon, \ee
where $\epsilon \equiv 1-\xi$ for small $\epsilon \ll 1$.  Therefore the faster the mirror, the colder the black hole.  The mirror must be fast enough for equilibrium to be achieved in the first place, so this result demonstrates a balancing act between warmth and equilibrium.    

\subsection{Life Expectancy}
If backreaction effects of the radiated energy on the geometry is ignored, then the black hole, as described using the usual tortoise coordinate, emits thermal radiation even at very late times, resulting in divergent energy. This is in clear contradiction with finite mass.  The inverse relationship between mass and temperature,
$$T = \frac{1}{8\pi M},$$
makes it explicitly clear that large mass black holes are colder than small mass black holes.  Interestingly, the evaporative stress-energy flux \cite{Good:2016atu}, $F(t)$, plateaus as $\xi \to 1$, corresponding to a time period of thermal emission for the black hole in a quasi-equilibrium state.  This increase in drift speed of the coasting mirror corresponds to its {\it{thermal lifetime}}, which increases, (see Fig.~(2) in \cite{Good:2016atu}), as $\xi \to 1$, but remains finite.  The physical picture is that particles are emitted during gravitational collapse -- in the initial phase the spectrum is far from thermal (some might refer to this as ``pre-Hawking radiation'', see \cite{1710.01533,Mann:2018jcf}), it can then approach thermality arbitrarily close for arbitrary long time, before finally deviating away again (one could say, ``post-Hawking radiation").


We note that the faster the mirror, the more massive the corresponding black hole and the longer its lifetime.  Strictly speaking, faster $\neq$ more massive; as $M$ and $\xi$ are two independent uncoupled parameters of the system.  However, faster $=$ more massive, under the condition $\xi \approx 1$ and the two previous assumptions: (1) $m\ll M$, and (2) $M = E(\xi)+m$.  Using the prototypical situation where $t_{\textrm{ev}} \propto M^3$ gives a lifetime that scales as $t_{\textrm{ev}} \propto \gamma^3$ from Eq.~(\ref{speedmass}). 
\section{Conclusion}

In this work, by utilizing the black hole - moving mirror correspondence \cite{Good:2016oey}, we have generalized the tortoise coordinate to the ``giant tortoise coordinate''. The resulting black hole gives rise to (1) finite evaporation energy, which is arguably a lot more physical, and  (2) overall unitarity is preserved, with  remnant remains after Hawking evaporation. 

We have extended the black hole radiance model to mimic black holes more accurately using a straightforward coordinate generalization of the usual spherically symmetric shell of pressureless massless matter, while remaining within the realm of tractable models (i.e. the beta Bogolubov coefficients have been solved in \cite{Good:2016atu}).
Future improvement on this line of extension would be a natural generalization that did not lead to drift but instead had the origin of coordinates come back to rest in the far future, i.e. a black mirror counterpart that was asymptotically static\footnote{For relevant asymptotically static investigations see \cite{Good:2017kjr} and references within.}.  This extension would be very interesting as it would avoid soft particle production and give total finite entropy production.

A missing piece of this current model is an explicit description of the non-trivial spatial curvature in the region where there is lack of continuity. Despite this, signatures of particle emission are manifest in the stress-energy\cite{Good:2015nja}.  It would be interesting, although ambitious, to find a consistent solution without the emission of negative energy flux. This would throw doubt on the need for a black hole ``death gasp'' \cite{Good:2015nja,1405.5235}.  Such a solution would benefit from being expressed within this framework and might be done using a further altered set of coordinates.

The crucial physical feature that enforces the late-time Hawking thermal emission result (but not post-Hawking radiation) is the very large redshift suffered by the wave that propagates from the proximity of the horizon to infinity encoded in the matching condition, Eq.~(\ref{m1}).  This is left unchanged by the giant tortoise coordinate, Eq.~(\ref{tortoise1}), as long as $\xi \approx 1$.  Moreover, the new matching condition, Eq.~(\ref{m2}), remains insensitive up to the unimportant $v_H$ constant, to the particular process of gravitational collapse.  The thermal emission in the case of the giant tortoise coordinate is still {\it{universal}} and does not depend on the nature of collapse. This behavior has counterpart in the emergent scale-independence that occurs for the proper acceleration, $\alpha = \tau^{-1}$, in the eternally thermal case \cite{Good:2017ddq}.

The evaporation process, in the usual tortoise case, evolves to thermal equilibrium, enduring perpetually, even at very-late-times, ultimately resulting in divergent temperature as mass drops to zero. While the effects of backreaction should be calculated before this temperature singularity occurs, and a primary consequence should be to render the total evaporation energy finite \cite{Fabbri:2005mw}, the giant tortoise provides total finite evaporation energy by utilizing an additional parameter, $\xi \approx 1$, which strictly keeps the origin of coordinates of the black hole at asymptotically sub-light speeds, $\xi < 1$, and rejects uncompromising continuity of the metric across the shock ray in order to preserve unitarity.
 
 Overall, the giant tortoise does not allow the black hole to emit thermal radiation at ultra-late times, rendering the total energy finite.  This is in clear agreement with the finite amount of energy contained inside the black hole.  

\section*{Acknowledgments}
MG thanks Paul Anderson, Xiong Chi, Eric Linder and Frank Wilczek for stimulating discussions. MG was funded in part from the Julian Schwinger Foundation under Grant 15-07-0000 and the ORAU and Social Policy grants at Nazarbayev University. MG also thanks the Center for Gravitation and Cosmology of Yangzhou University for hospitability during his visit. 
YCO acknowledges grant No.11705162 of  National Natural Science Foundation of China, as well as  grant No.17Z102060070 of China Postdoctoral Science Foundation Fund.  

\appendix

\end{document}